# VALIDATION OF HOMOGENIZED FINITE ELEMENT MODELS OF HUMAN METASTATIC VERTEBRAE USING DIGITAL VOLUME CORRELATION

Chiara Garavelli[1,2], Alessandra Aldieri[1,2], Marco Palanca[2], Enrico Dall'Ara[3,4], Marco Viceconti[1,2]

[1] Medical Technology Lab, IRCCS Istituto Ortopedico Rizzoli, Bologna (IT)

[2] Department of Industrial Engineering, Alma Mater Studiorum - University of Bologna (IT)

[3] Department of Oncology and Metabolism, Mellanby Centre for Bone Research, University of Sheffield, Sheffield, United Kingdom

[4] INSIGNEO Institute for in Silico Medicine, University of Sheffield, Sheffield, United Kingdom

**ORCID IDs**

| | |
|---|---|
| Chiara Garavelli | https://orcid.org/0000-0002-6921-0730 |
| Alessandra Aldieri | https://orcid.org/0000-0002-2397-3353 |
| Marco Palanca | https://orcid.org/0000-0002-1231-2728 |
| Enrico Dall'Ara | https://orcid.org/0000-0003-1471-5077 |
| Marco Viceconti | https://orcid.org/0000-0002-2293-1530 |

## ABSTRACT

The incidence of vertebral fragility fracture is increased by the presence of preexisting pathologies such as metastatic disease. Computational tools could support the fracture prediction and consequently the decision of the best medical treatment. Anyway, validation is required to use these tools in clinical practice. To address this necessity, in this study subject-specific homogenized finite element models of single vertebrae were generated from µCT images for both healthy and metastatic vertebrae and validated against experimental data. More in detail, spine segments were tested under compression and imaged with µCT. The displacements field could be extracted for each vertebra singularly using the digital volume correlation full-field technique. Homogenised finite element models of each vertebra could hence be built from the µCT images, applying boundary conditions consistent with the experimental displacements at the endplates. Numerical and experimental displacements and strains fields were eventually compared. In addition, because also clinical CT scans were available for the same specimens, the outcomes of a µCT based homogenised model were compared to the ones of a clinical-CT based model. Good agreement between experimental and computational displacement fields, both for healthy ($R^2$ = 0.69÷0.83, RMSE% = 3÷22%, max error < 45 µm) and metastatic ($R^2$ = 0.64÷0.93, RMSE% = 5÷18%, max error < 54 µm) vertebrae, was found. The comparison between µCT based and clinical-CT based outcomes showed strong correlations ($R^2$ = 0.99, RMSE% < 1.3%, max error = 6 µm). Furthermore, the models were able to qualitatively identify the regions which experimentally showed the highest strain concentration. In conclusion, the combination of the experimental full-field technique and the in-silico modelling allowed the development of a promising pipeline for the validation of fracture risk predictors, although further improvements in both fields are needed to better analyse quantitatively the post-yield behaviour of the vertebra.

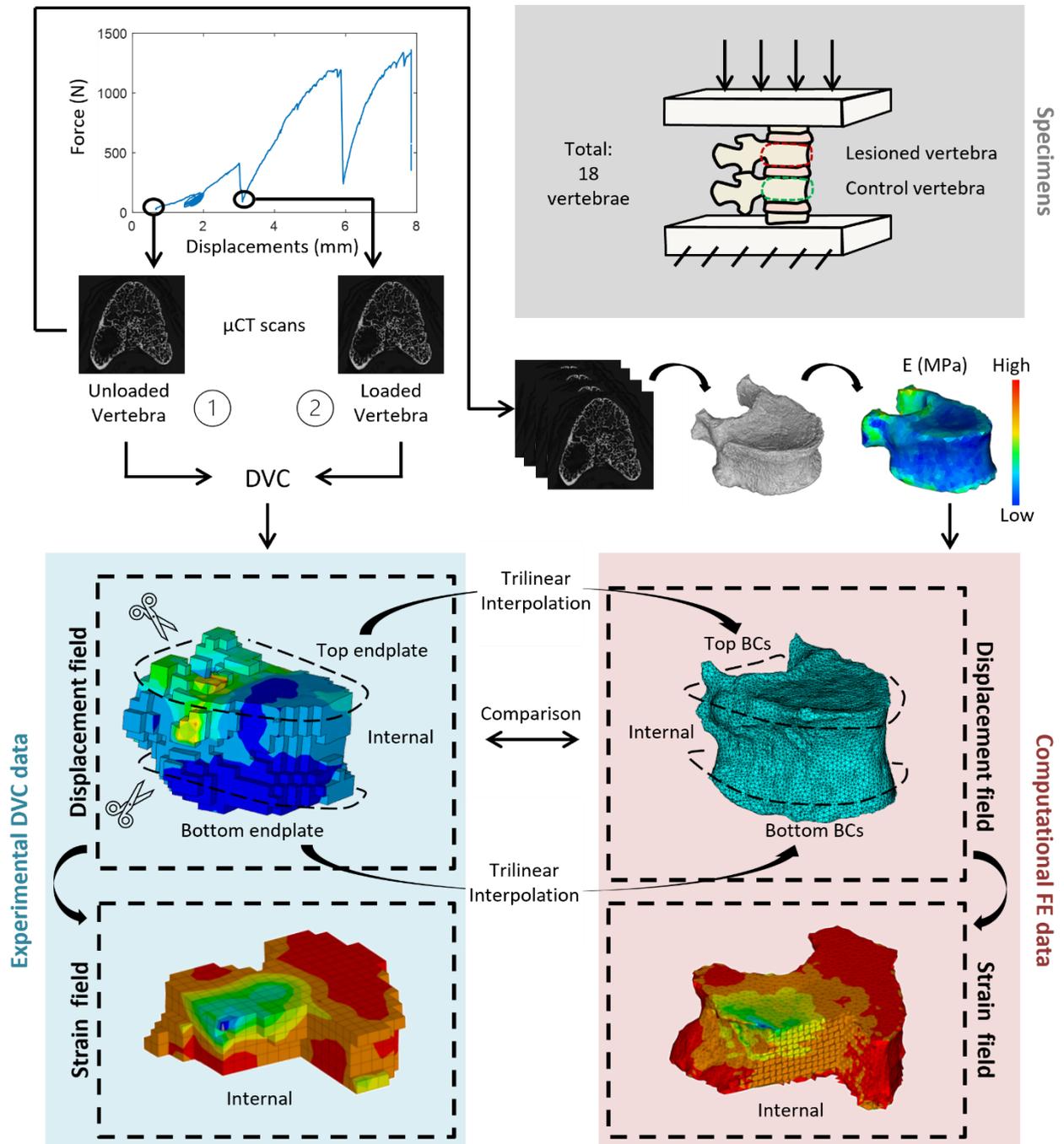

**Graphical abstract**

# INTRODUCTION

In 2020 the World Health Organization (WHO) reported an incidence of more than 18 million cancer occurrences worldwide and presented a prevision of almost 20 million for the year 2040 [1]. Among them, one-third of the patients presented signs of spinal metastasis [2]. The presence of such metastases, especially lytic, has been demonstrated to increase the vertebral risk of fracture [3]. Although the incidence of this type of metastatic lesions can appear moderate, the comorbidities connected to the fracture events at the vertebrae should be considered [4], as well as the associated decrease in the quality of life (Alexandru et al., 2012). In light of this, constant monitoring of vertebral stability in pathological patients is crucial, aiming to prevent the fracture from occurring.

The finite element (FE) methodology has been successfully adopted to evaluate the stability of vertebrae with lytic lesions [6] showing the ability to provide information in cases where the Spinal Instability Neoplastic Score lacked in specificity [7]. A similar strategy has also been applied to vertebral bone affected by osteolytic lesions, allowing to obtain better assessment of bone fragility compared to volumetric bone mineral density, bone volume fraction and trabecular separation measurements [8]. In addition, it has been shown that metastatic vertebrae strengths from experimental tests and from FE models correlated well ($R^2 = 0.78$) regardless of the lesion phenotype (Stadelmann et al., 2020). However, no direct validation of displacements and strains fields predicted by the models has been presented in these studies. Moreover, they investigated simplified boundary conditions only.

In order to fill this gap, Digital Volume Correlation (DVC) represents a useful tool because it provides full-filed information over the whole bone volume [10], has been widely used to validate bone models [11] and can provide experimentally matched boundary conditions (BCs) for the models [12], [13]. Vertebral body FE models have already been validated using DVC both at the tissue level, i.e., µFE models validation [14], [15] and at the organ one, i.e., hFE models validation [16]. Moreover, DVC technique has already been applied to validate vertebral FE models where the load was transmitted through the discs [17]. However, in that study the focus was on osteoporotic vertebrae and the validation was carried out by averaging the results within areas of interest where the vertebra had been divided.

The aim of this work was thus to develop µCT-based homogenized finite element (hereinafter referred as hFE) models, for both healthy and metastatic vertebrae, and to validate their prediction in terms of displacements and reaction forces against experimental data. DVC measurements were used to perform the validation. Given that µCT are not performed in clinics, the outcomes of one µCT-based hFE model were also compared to the ones of clinical-based hFE model of the same vertebra. Eventually, the ability of the hFE models to highlight the regions with higher strain concentrations were assessed.

# MATERIALS AND METHODS

The following pipeline was applied on 18 vertebrae, divided as follows: 10 healthy and 8 metastatic, specifically 4 lytic and 4 mixed. The comparison between hFE models generated from different resolution scales (clinical CT and µCT) was performed on a single healthy vertebra.

**Mechanical Testing**

Experimental analyses were conducted on thoracolumbar cadaveric specimens obtained from an ethically approved donation program (Anatomic Gift Foundation, Inc.). Each specimen was composed of four vertebrae: the most cranial and the most caudal ones were embedded in the polymethylmethacrylate and were used to apply the load while the two (one showing signs of lytic or mixed metastatic lesions and one healthy control) in between were free to move through the action of the intervertebral discs. During the preparation of the specimen, posterior elements were removed (Fig.1A). The specimen was mounted on a jig and inserted into a µCT scanner where an unloaded scan (Fig.1C) was acquired with 39 µm isotropic voxel size (current: 114mA, voltage: 70kVp, integration time: 300ms, power: 8W). The mechanical test was performed as follows and as described in greater detail in Palanca et al., 2023: 1) a compressive load inducing physiological strains on the surface of the healthy vertebra [19] was applied at the cranial vertebra; 2) the load was increased of three times; 3) the load was increased up to failure (Fig.1B). After each load step a µCT scan was acquired following relaxation (Fig.1D). DVC, implemented through the BoneDVC algorithm [20], was later performed on each combination of the unloaded scan and one of the loaded scans for the same specimen. Due to the high computational cost required by the BoneDVC algorithm, the original images were cropped, before to start the elastic registration, in order to contain only one vertebra at a time. DVC was carried out based on a 1.95 mm sized grid, where displacements values were thus available (Fig.2A). The displacement field was imported into a FE analysis environment (ANSYS Inc., Mechanical APDL), where the strain field was obtained through differentiation. Further technical details about the DVC technique can be found in [18].

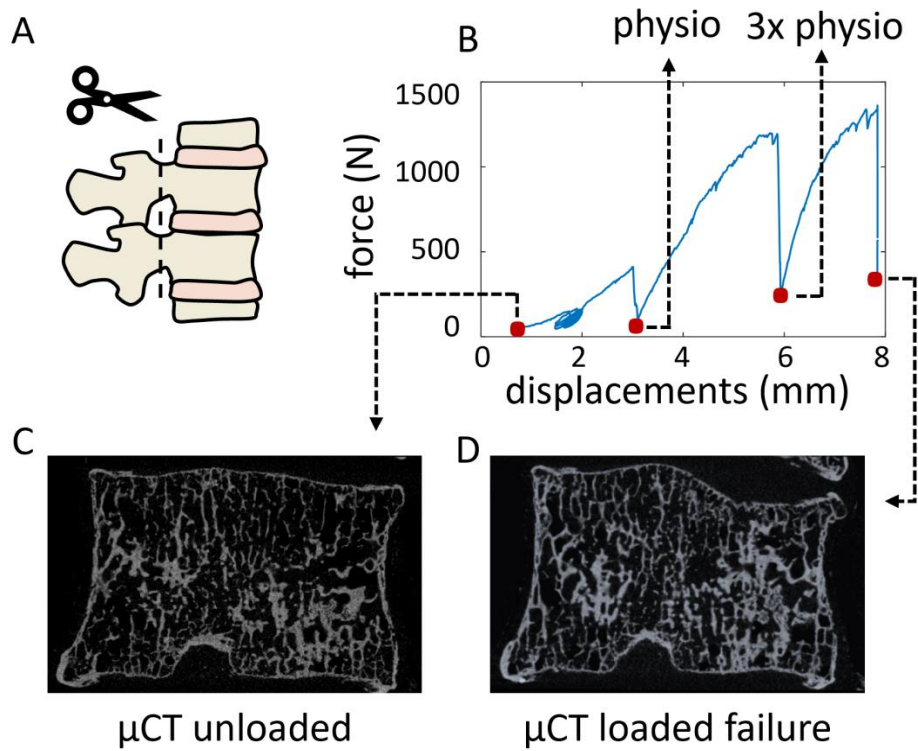

**Fig.1**: Experimental test. The specimens were composed by four vertebrae, but the two extreme ones were used to transmit the load only, and the posterior elements were removed (A). During the compression test (B), μCT scans were acquired at the unloaded condition (C), then after each load step (red dots), until failure (D).

**Finite Element Analysis**

In analogy with the experimental analysis, where the DVC measurements were carried out on one vertebra at a time, FE models of each vertebra were also created independently. Moreover, both the hFE and the μFE models were generated from the μCT images. In this way, FE models and DVC data shared the same reference system, allowing to reduce errors connected to the registration process. The possibility to generate an hFE model based on a μCT scan has been presented in several works [21]–[23]. Additionally, the outcomes of one μCT-based model were compared to the outcome of the same model mapped on a clinical CT to assess the correctness of the procedure. The steps required for the two kinds of model are reported in the following two paragraphs.

The hFE model was created based on the unloaded μCT scans. First, the μCT images were segmented adopting a semi-automatic segmentation procedure (Materialise Mimics), and a 10-node tetrahedral structural solid mesh was later created (ICEM, ANSYS Inc.), with an edge length equal to 1 mm [6]. The material properties were assigned analogously for the healthy and pathological vertebrae. μCT grey levels were calibrated to obtain density values, using hydroxyapatite phantoms. Then, Bonemat software developed at Rizzoli Orthopaedic Institute was used to integrate the voxel density over each element of the mesh. Coefficients from the literature were used to define the density-elasticity relationship [24]. Subsequently, boundary conditions were assigned to the hFE model in order to reproduce the experiments. For each vertebra, the two DVC grid slices falling inside the bone and closest to the endplates (upperBC and lowerBC) were selected. The displacements of upperBC and lowerBC were interpolated onto the FE nodes at the extremity of the vertebral body through a trilinear interpolation algorithm (Matlab, Mathworks Inc.), such that the FE model could be loaded in displacement using experimental data (Fig.2B).

For one specimen, the same hFE tetrahedral mesh already mapped on μCT was roto-translated to a clinical CT acquired on the same specimen for a previous study [19] using a procedure accurately described elsewhere [25]. Then, the hFE was mapped on the clinical CT scans, calibrated using ESP phantom. A correction of the calibration was also applied (Schileo et al., 2008) and eventually a conversion from element density to elasticity [24]. After that, the inverse transformation matrix was applied to the hFE nodes to reposition the model back onto the μCT reference system. Eventually, the model was solved by applying the same boundary condition previously described for the μCT-based model.

The hFE simulations were run on a FE analysis environment (ANSYS Inc., Mechanical APDL) and performed on a local computer (parallel distributed memory over 6 cores with 64 GB of RAM (Intel(R) Xeon(R) E-2276G CPU 3.80GHz) and required around 1 hour.

Two additional tests were performed on one vertebra only, and are reported in Appendix A. The tests are respectively, the inclusion of an elastoplastic behaviour to the hFE model, and the propagation of the error on displacement to an error on the strains. The addition of plasticity is intended to improve the prediction of deformation in vertebrae that are nearing their elasticity limit. This is particularly useful for predicting fracture risk because once the limit is exceeded, the vertebra is permanently damaged, and we only need to know that it has fractured, not the exact value of the deformation. What's interesting is the moment of damage accumulation in the area just before the fracture. That's why it was important to determine whether adding plasticity would help us better track this phenomenon. The study of error propagation is based on the fact that it isn't possible to use the exact value predicted by the DVC to validate the strain of the models. However, the fracture definition is based on the strain. Therefore, we decided to quantitatively validate the displacements and qualitatively validate the strains. Then, we defined a procedure to evaluate the error propagation from displacements to strains.

**Validation and Comparison**

In order to compare the hFE model against experimental data the hFE nodal displacements were interpolated at the DVC points location using the shape functions of the elements. Specifically, a two-fold validation procedure was followed: at first, all DVC point locations were used, secondly, only a subsample of the aforementioned DVC points, namely the one falling inside a trabecula (Fig.2D) or within the cortical shell. The point-to-point validation was always performed on the central 75% of the modelled vertebra, in order to be sufficiently far from the nodes used to apply the boundary conditions (Fig.2C).

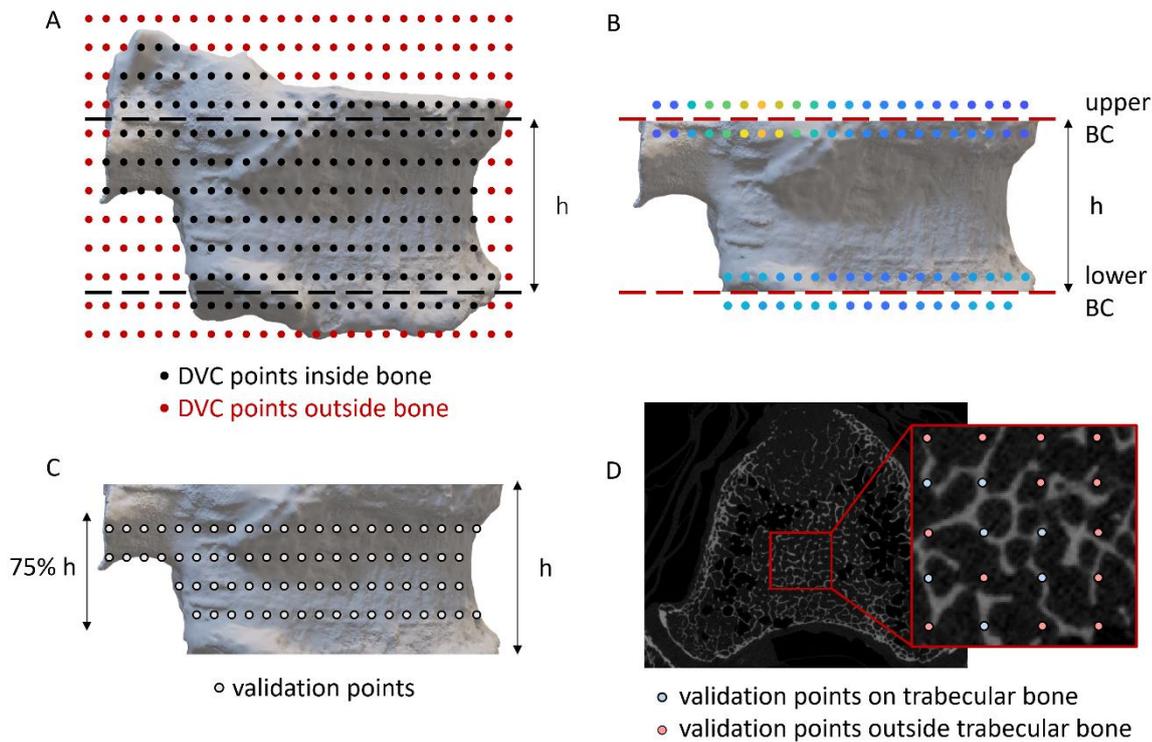

**Fig.2**: Definition of the DVC points (A, only red dots provide usable information) used for BCs assignment (B) and validation of the models (C). Among the latter only the validation points falling inside a trabecular bone were used also in the comparison between hFE and µFE (D).

**Metrics**

The agreement between hFE and DVC displacements was reported by computing the intercept and the slope of the linear regression, the coefficient of determination ($R^2$), the root mean squared error (RMSE), and the percentage RMSE (RMSE%), obtained normalizing the RMSE by the higher experimental displacement. Maximum and minimum principal strains obtained from the hFE model have been qualitatively compared to the one provided by the experimental test, to compare the location of the regions in which the deformations were concentrated.

In addition, the agreement between the axial reaction forces predicted by the FE analysis and the ones measured during the experimental tests was assessed, in term of $R^2$, RMSE% and maximum error. The numerical reaction forces were computed in the axial direction at the nodes of BClineDOWN.

**Exclusion Criteria**

Among the tested specimens, some vertebrae needed to be excluded from the validation process. The following exclusion criteria were used: (i) more than 25% of the correlation points present experimental deformations over Bayraktar failure limits [27], (ii) the lytic metastatic lesion has destroyed the trabecular lattice of the vertebra, for more than half of its body, and/or (iii) the prediction errors strongly correlate with the DVC uncertainties at the same locations. Each vertebra matching at least one of the criteria was excluded, while the other one from the same specimen was validated as usual.

## RESULTS

The comparison between µCT-based hFE and DVC displacements is shown in Fig.3, displayed differently according to the health status of the vertebra (control, lytic metastasis, mixed metastasis) they originated from. It is possible to observe that for all the directions and for all the categories the model slightly underestimated the experimental values. Anyway, no specific dependence of the goodness of fit on the health status of the vertebra was found. In fact, the strongest correlation in the anterior-posterior direction was found for the lytic vertebrae ($R^2 = 0.93$), while in the cranio-caudal direction that was found for the healthy ones ($R^2 = 0.83$). The RMSE% for the control vertebrae were in the range 3-22% and the maximum error was lower than 45 µm, while for lesioned vertebrae RMSE% was in the range 5-18% and the maximum error was lower than 54 µm. For pooled data the RMSE% were 6% for anterior-posterior and 13% for the cranio-caudal. The correlation indexes of each vertebra are also reported in the boxplots in Fig.4. The entity of the displacements in the right-left direction resulted to be lower than the voxel size for most of the points. For this reason, these displacements were considered too close to the experimental measurements uncertainties and were therefore excluded from the validation process.

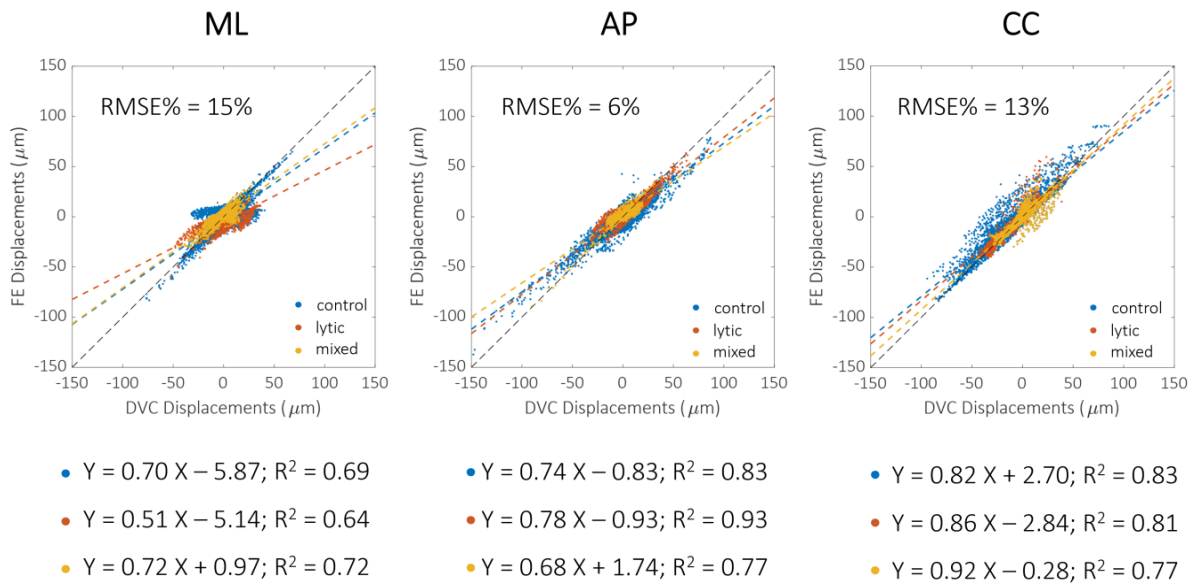

**Fig.3**: Comparison between DVC (horizontal axis) and hFE (vertical axis) displacements, on mediolateral (ML), anteroposterior (AP) and craniocaudal (CC) directions respectively. Control vertebrae are reported in blue, lytic in orange and mixed in yellow. For each type of vertebra regression lines and $R^2$ are also reported, while RMSE% in reported for the pooled groups.

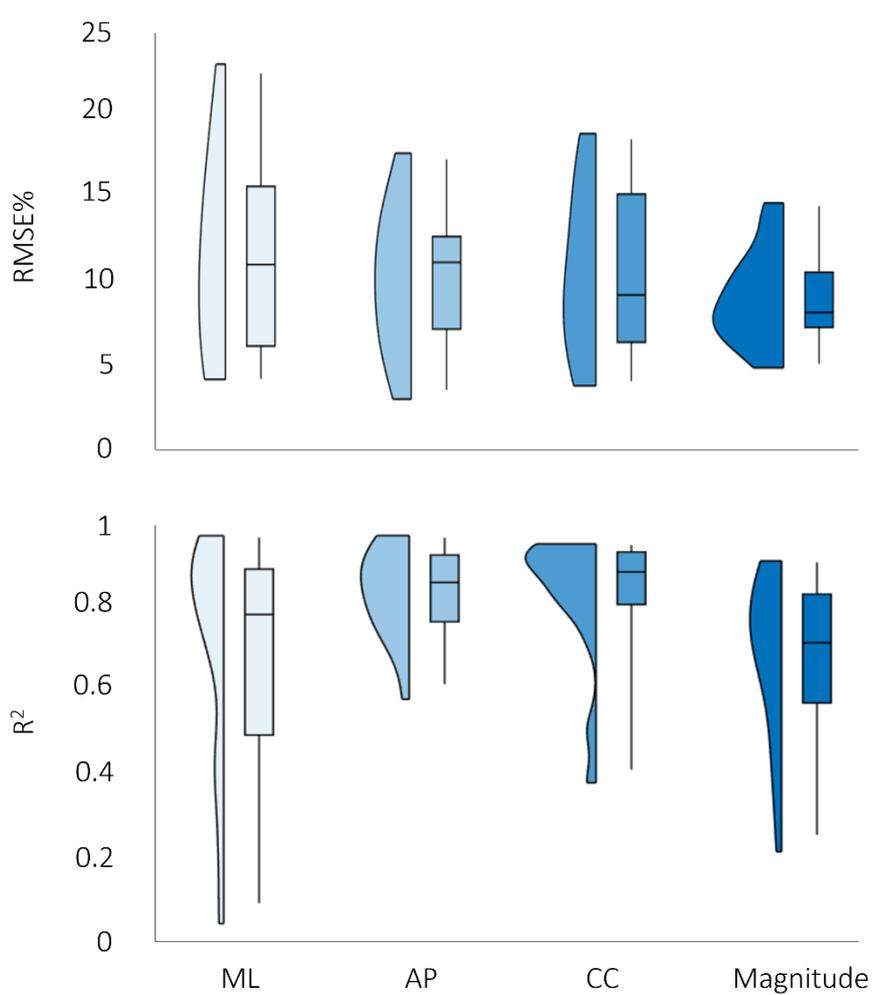

**Fig.4**: Boxplots reporting RMSE% and $R^2$ for all the vertebrae analysed. Red horizontal lines represent the ideal value for each index. Mediolateral (ML), anteroposterior (AP) and craniocaudal (CC) directions as well as the magnitude are reported.

Narrowing the analysis to those DVC points whose position lies within the trabecular bone, a substantial decrease in RMSE% is observable for the craniocaudal direction, while no strong differences are observed for the anteroposterior direction (Fig.5).

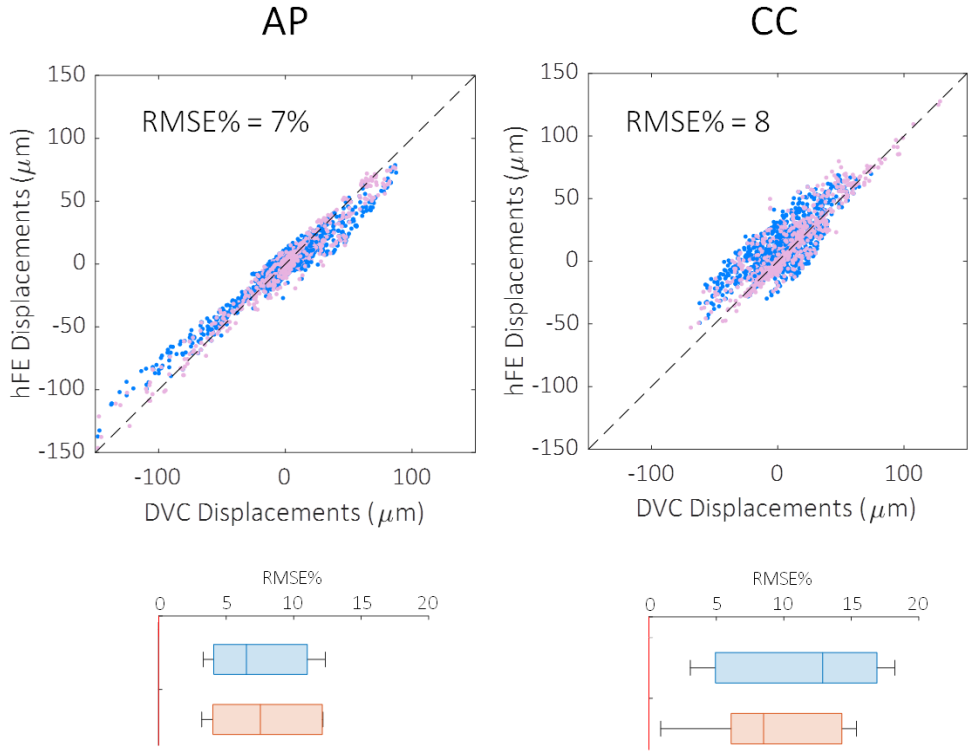

**Fig.5**: Scatter plots between DVC (horizontal axis) and hFE (vertical axis) displacements on trabecular locations only, on anteroposterior (AP) and craniocaudal (CC) directions respectively. Below each direction, the relative changes in RMSE% due to the restriction of the analysed points are reported (light blue before, pink after the restriction).

Furthermore, strong agreement in the displacements field was found between the μCT-based and the clinical-CT-based models, with RMSE lower than 0.75 μm (RMSE% < 1.3%) and $R^2$ higher than 0.99. The maximum difference among all the nodes and considering the three cartesian directions was 6 μm (Fig.6). In terms of predicted reaction forces, the difference between μCT-based and clinical-CT-based models is around 1%, specifically 2088 N in the former case and 2119 N in the latter.

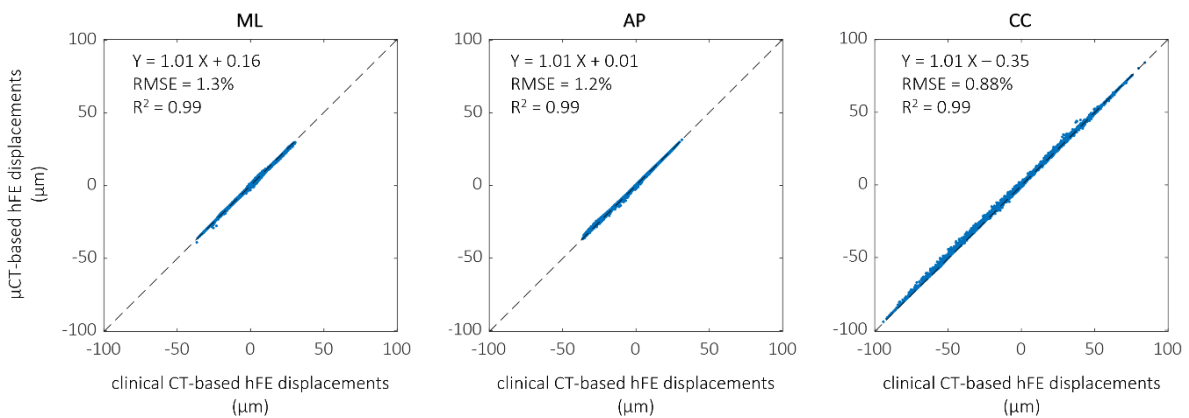

**Fig.6**: Correlations between displacement fields predicted by clinical-CT- (horizontal axis) and μCT-based (vertical axis) models. Regression lines, $R^2$ and RMSE% are reported.

A qualitative comparison of experimental and simulated strains is also reported. The distribution of maximum and minimum principal strains highlighted that the hFE models were able to select the regions of the vertebral body providing the highest strain concentration (Fig.7).

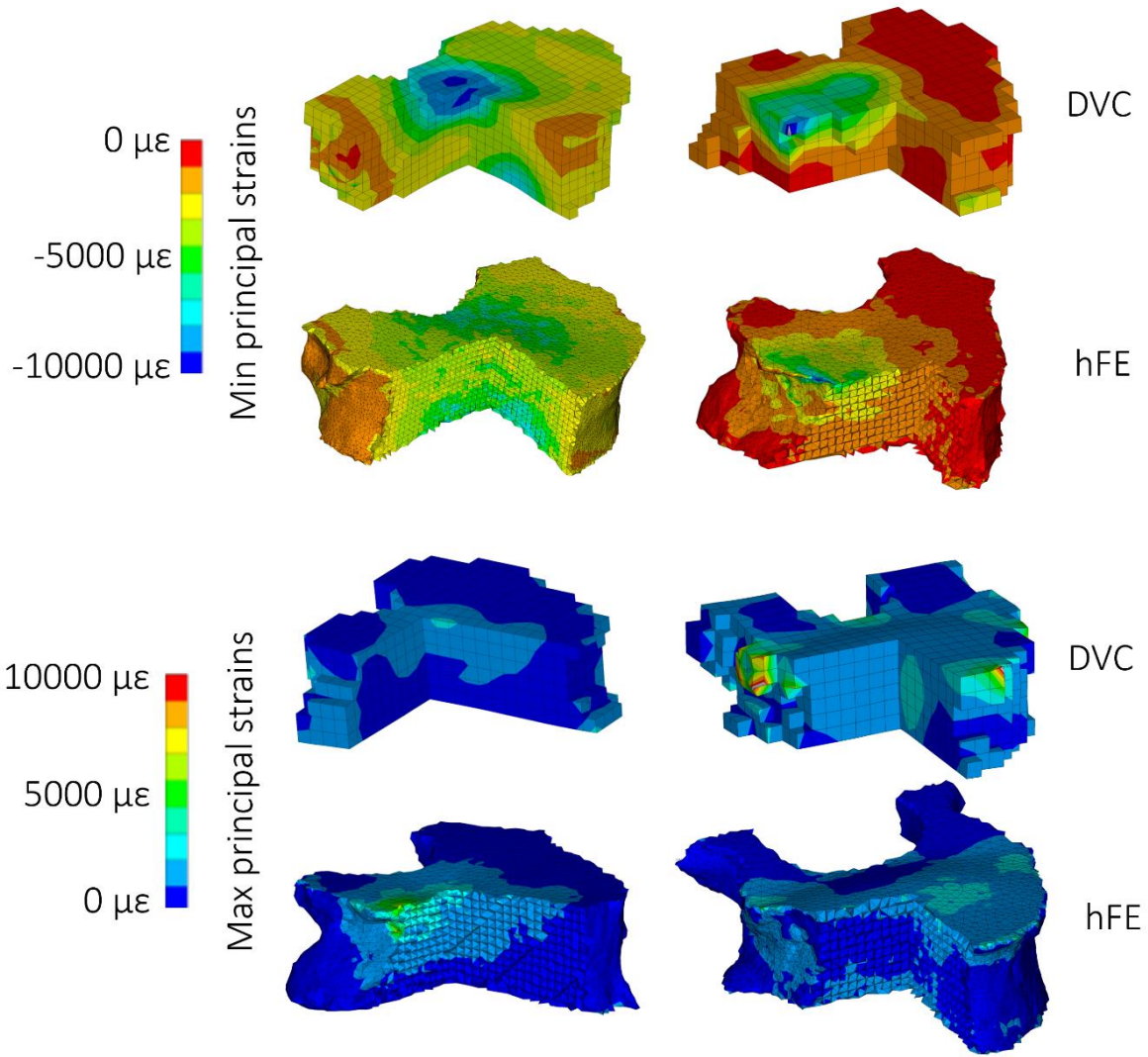

**Fig.7**: Qualitative comparison between DVC and hFE strains. Four different vertebrae are reported as examples; for each one the upper contour plot refers to the DVC hexahedral grid, while the lower refers to the hFE tetrahedral mesh. The figure highlights that for the majority of the specimens the regions presenting higher deformations are correctly predicted (B, C). However, there are still some specimens for whom the FE models fail in founding the highly deformed regions (A, D).

Lastly, quite good correlation was found between the axial reaction forces predicted by the FE analysis and the ones measured during the experimental tests, with $R^2 = 0.65$, RMSE = 19% and maximum error equal to 1368 N (Fig.8).

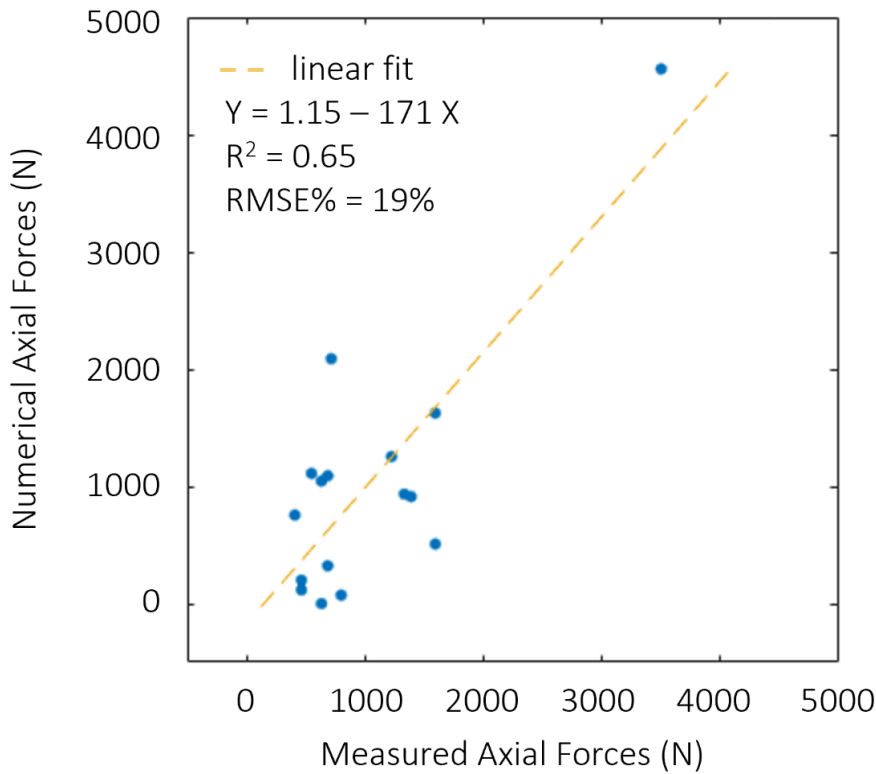

**Fig.8**: Comparison of axial reaction forces predicted by the vertebral hFE models and the experimentally measured ones. Regression line (yellow), $R^2$ and RMSE% are reported.

Among the analysed vertebrae, three (1 control, 1 lytic and 1 mixed) were excluded from the validation because they fitted one of the Exclusion Criteria (Appendix B).

## DISCUSSION

This study aimed to assess the accuracy of subject-specific hFE models of the vertebra with experimentally matched BCs directly applied in the prediction of displacements and strains compared to a widely used experimental method.

Good agreement between DVC and hFE displacement fields, both for healthy ($R^2 = 0.69 \div 0.83$, RMSE% = $3 \div 22\%$, max error < 45 μm) and metastatic ($R^2 = 0.64 \div 0.93$, RMSE% = $5 \div 18\%$, max error < 54 μm) vertebrae, was found. Furthermore, the models were able to qualitatively identify the regions which experimentally showed the highest strain concentration. The derivation of displacements error to strains errors is also reported in Appendix A, with average errors around three thousands of microstrains. However, it is important to highlight that the errors committed by the homogenised FE models were obtained in presence of a DVC uncertainties identified in zero-strain conditions of 6÷9 μm for the displacements field and average errors around two thousands of microstrains for the strains field [28]. Consequently, the difference between the prediction errors and the experimental uncertainties were less than two order of magnitude. This does not allow to consider the model validated however the theories embodied by the models have resisted attempts at falsification by the DVC experimental. In addition, quite satisfactory agreement was found in the reaction forces ($R^2 = 0.65$, RMSE% = 19%) corroborating the density-elasticity law employed.

The presented results in terms of FE displacement prediction of DVC measurements were found to be in agreement with the ones reported by Palanca et al., 2022. There, μFE models of porcine vertebrae were developed and experimental derived displacements were applied as BCs, supporting the implementation performed in this study, to be eventually compared against DVC (RMSE% =1.01-14.51, $R^2$ = 0.65-1.00, slope = 0.77-1.19). The lower correlation found in this study compared to Palanca et al., 2022 could be explained considering that the work used porcine healthy specimens with induced lesions, while this study tested human vertebrae, some of them with lytic or mixed metastasis, and this aspect increased the difficulty in the material modelling. Jackman et al., 2016 performed human vertebra FE models validation against DVC with an experimental set-up similar to the one in this study. Their findings for the compression tests with experimentally matched boundary conditions showed a median error in displacements in the range of 20-80% (average = 49%). Computing the same value for the presented data the resulting range is 13-109% (average = 39%), highlighting a comparable fitting of the experimental data. DVC techniques have also been applied in the validation of hFE models of the scapular bone, analysing both the displacements [29] and the strains fields (Kusins et al., 2020) respectively with point-to-point and averaged comparisons. This study achieved slightly inferior results in terms of displacement field prediction for the same boundary condition application approach ($R^2$ =0.40-0.98, slope =0.43-1.32 in this study versus $R^2$ = 0.79-1.00, slope = 0.87-1.09 in Kusins et al., 2019). Also, the correct identification of the regions which experimentally showed the highest strains concentration by the hFE was confirmed in Kusins et al., 2020.

It is necessary to highlight some limitations of this work. Firstly, as explained above, the comparison was made only for vertebrae that did not show visible signs of failure at the time of the loading scan. This is because the developed models were linear elastic and, therefore intrinsically unable to correctly predict deformations beyond the elastic regime. Further improvements in both computational and experimental methods are needed to quantitatively analyse the post-yield behaviour of the vertebra. Secondly, the necessity to sacrifice the endplates regions to inform the models excluded these regions from the validation, although they are known to be of major interest for the failure mechanisms [18]. This criticality could be overcome only when it will be possible to obtain accurate DVC measurements for soft tissues (and therefore also of the discs), to be able to move the application area of the boundary conditions externally to the endplates.

In conclusion, the combination of the experimental DVC technique and the FE modelling technique has allowed to develop a promising pipeline for validation of in silico predictors of fracture risk. However, as already mentioned, the models prediction errors identified were not sensitively higher than the experimental data uncertainties, which prevented to consider the model validation successful. Two possible approaches could be followed to overcome this conundrum. On one side, to the use of imaging techniques able to reach a higher resolution, lower than a micrometre. Nevertheless, imaging tools able to achieve this level of resolution cannot currently be applied to a whole vertebra, but only to a portion of it, acquiring only some trabeculae at a time. On the other side, an extensive tests campaign where the numerical failure load would be validated against experimental data could be carried out, to provide evidence of the accuracy of hFE models on a different level.

# SUPPLEMENTARY MATERIAL

# Appendix A

The following two sections report tests performed on one vertebra only. The first aimed at assessing if the inclusion of a simple bilinear plastic behaviour to the hFE model can improve the models accuracy. The second aimed at defining a way to propagate the errors on displacement to errors on the strains.

**Plasticity**

An elastoplastic behaviour was also integrated in the hFE model development using a bilinear, isotropic behaviour, and a symmetric yield stress criterion. Similar approach have already been applied in the estimation of the mechanical competence of vertebrae with lytic metastases [6]. The plasticity coefficients were taken from the literature and were defined as follows.

$\sigma_{y1} = 21.70 * \rho_{app}^{1.52}$ [g/cm3]     [30]          (Eq.2)

$Ep_y = 0.05 \times E$  [MPa]          [24], [27]          (Eq.3)

The elastoplastic hFE displacement field was compared to the elastic hFE displacement field, as well as to the experimental measurements.

Elastic and elastoplastic hFE models showed good accordance in the craniocaudal direction (RMSE% = 2%), while lower accordance in the anterior-posterior one (RMSE% = 10%), and looking at the validation against the DVC measurement, it was possible to highlight a strong decrease in the data fitting (Fig.2s).

Since the correlations instead of improving worsen slightly, it was concluded that non-inclusion of the plasticity in the model was not the main source of the error in the prediction of the experimental outcomes.

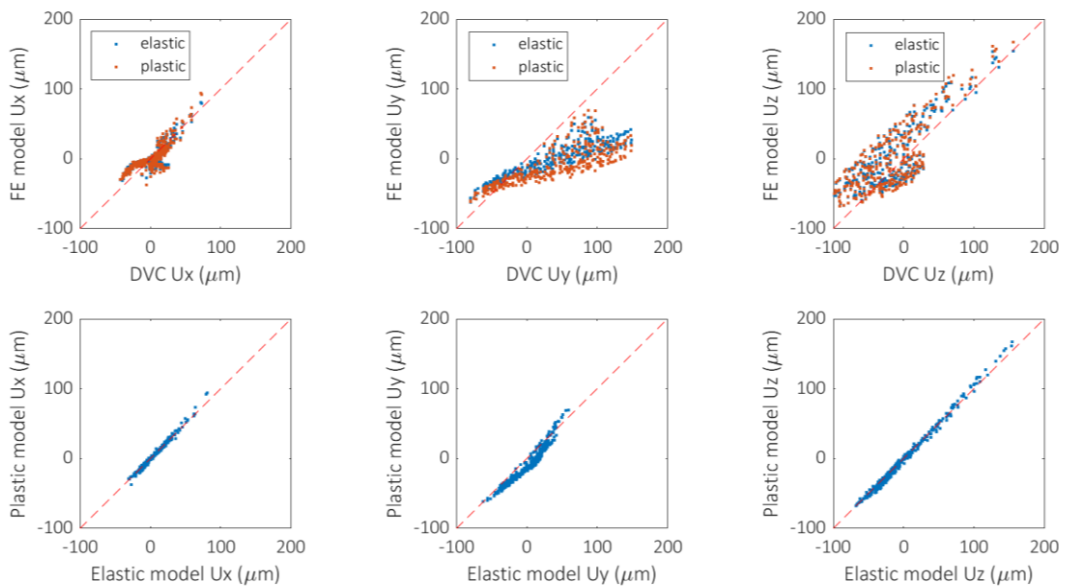

**Fig.2s**: In the first row, the correlations between displacement fields calculated by DVC (horizontal axis) and predicted by the FE (vertical axis) models, both with elastic (blue) and elastoplastic (orange) material properties assigned. In the second row, the correlations between the two models are also reported.

**Error Propagation**

The possibility of propagating the error on displacement to an error on strain was analysed, considering that the displacement is the original variable computed by the experimental procedure. Firstly, an estimation of the error on strain was performed considering the predictive error on displacement given by the difference between FE and DVC displacements and dividing it by the grid nodal spacing (~ 2 mm). The order of the error was compared to the order of the uncertainty on strains. Then, the point-to-point difference between hFE and DVC displacements at the DVC points locations was superimposed to each node of the hexahedral DVC mesh and the FE simulation environment (ANSYS, Inc.) was used to derivate the local errors on the strain field. Subsequently, the zero-strain value at the relative point was used to reduce the error on strains, to remove the portion of error explained by the experimental uncertainty, and the distribution of the residual error was examined.

Considering the correlations obtained for the displacement field, the estimation of the error on the strains resulted in the range 0.001-0.025 $\varepsilon$ while the experimental zero-strain were in the order of 0.001 $\varepsilon$, at worst an order of magnitude less. These results were confirmed also analysing the point-to-point distribution of the error on strains (Fig.2s). The comparison between the distribution of the error on the strains and the distribution of the experimental uncertainty for the same vertebra is also reported in Fig.3s.

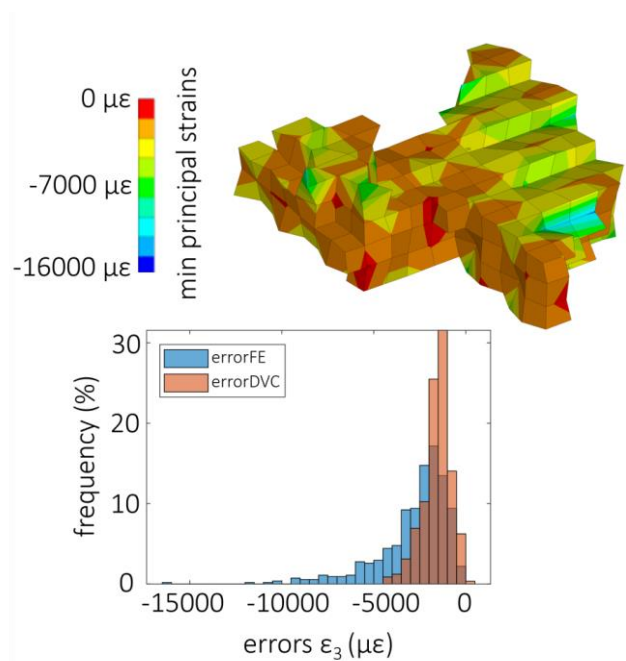

**Fig.3s**: On the top, the contour plot of the minimum principal strain due to the derivation of the displacement error is presented. On the bottom, the histogram of the predictive error (errorFE in blue) on the minimum

principal strain is compared to the one of the experimental uncertainties (errorDVC in pink) for the same vertebra.

## Appendix B: Excluded specimens

Among the analysed vertebrae, three have been excluded from the validation because they fitted one of the Exclusion Criteria, as reported below:

1) One vertebra showed experimental deformations over Bayraktar failure limits in more than 20% of the correlation points (Fig.4s).

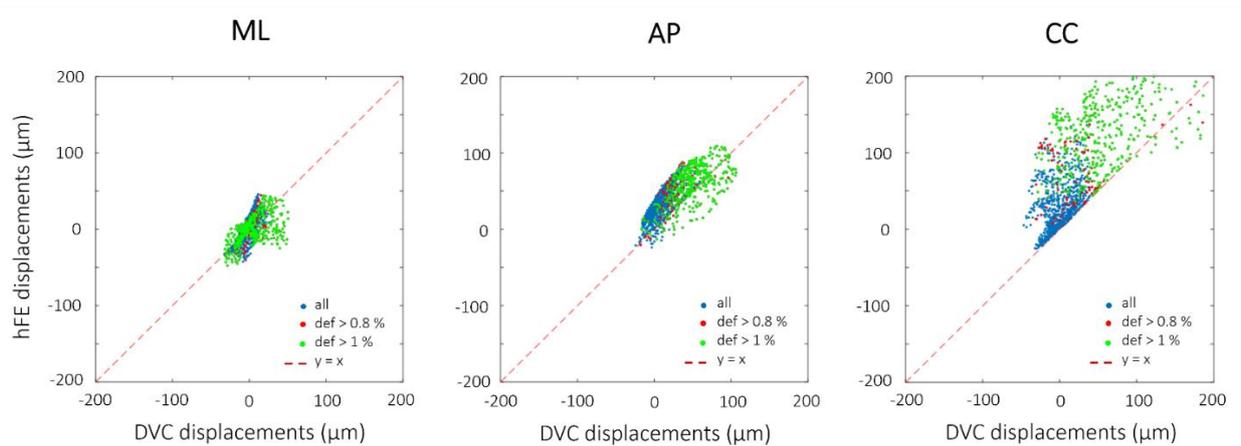

**Fig.4s**: Scatter plots between DVC (horizontal axis) and hFE (vertical axis) displacements, on mediolateral (ML), anteroposterior (AP) and craniocaudal (CC) directions respectively. The points in which the maximum and/or the minimum principal experimental strains overcome 0.8% (red) and then 1% (green) are highlighted in order to show that in the points with higher deformations the model have higher difficulties in reproducing the experimental data. The points in which all the strain components are lower than 0.8% are reported in blue.

2) One vertebra had a lytic metastatic lesion that had destroyed the trabecular lattice for more than half of its body. This strongly reduced the points of the DVC grid inside the bone (Fig.5s).

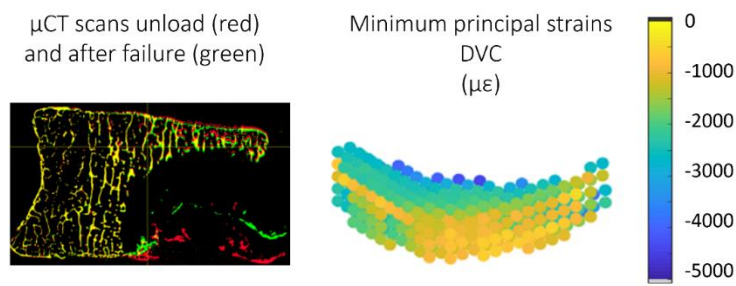

**Fig.5s**: The lytic lesion in this vertebra was bigger than half of the vertebral body as is possible to see from the µCT scan on the left. On the right the DVC points able to correlate are reported.

3) One vertebra showed a strong dependence of the prediction errors to the DVC uncertainties at the same locations (Fig.6s). Predictive error was computed has the absolute difference between DVC and FE displacements at each correlation point. DVC errors at zero-strain conditions were computed as the average of the six components of the strains derived from the experimental displacements computed applying BoneDVC algorithm to two unloaded scans of the same vertebra [28].

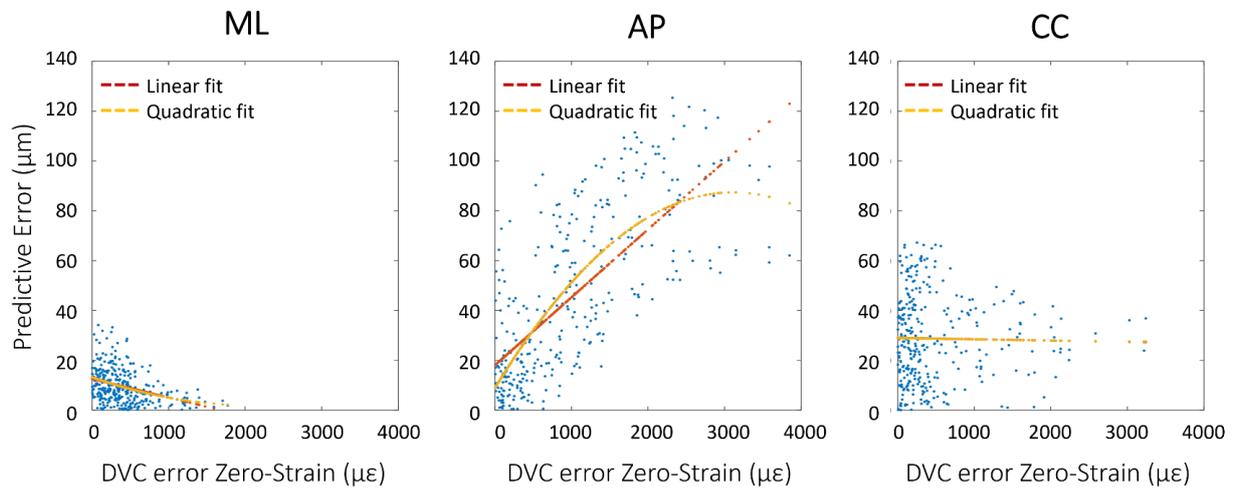

**Fig.6s**: Scatter plots between DVC uncertainties (horizontal axis) and predictive errors on displacements (vertical axis), on mediolateral (ML), anteroposterior (AP) and craniocaudal (CC) directions respectively. In the AP direction strong dependence was highlighted.